\documentclass[11pt]{article}
\usepackage[utf8]{inputenc}

\usepackage[colorlinks=true, citecolor=blue]{hyperref}
\usepackage{amsmath,amssymb,amsthm}
\usepackage{mathtools}
\usepackage{graphicx}
\usepackage[font=small,labelfont=bf,width=\textwidth]{caption}
\usepackage{subcaption}
\usepackage{fullpage}
\usepackage{color}
\usepackage[usenames,dvipsnames]{xcolor}
\usepackage{tikz}

\usepackage{fullpage}

\RequirePackage[sc]{mathpazo}
\RequirePackage[T1]{fontenc}

\usepackage[capitalize,nameinlink,noabbrev]{cleveref}

\usepackage[maxbibnames=99,sorting=nyt, giveninits=true]{biblatex}
\renewbibmacro{in:}{. In:}
\theoremstyle{remark}

\usepackage{authblk}

\title{Reductions of nonlocal nonlinear Schr\"odinger equations to Painlev\'{e} type functions}

\author{Jonathon Liu\thanks{\href{mailto:jliu8656@alumni.sydney.edu.au}{jliu8656@alumni.sydney.edu.au}}}
\affil{School of Mathematics and Statistics, University of Sydney}
\begin{document}

\maketitle

\begin{abstract}
In this paper, we take ODE reductions of the general nonlinear Schr\"odinger equation (NLS) AKNS system, and reduce them to Painlev\'e type equations.
Specifically, the stationary solution is solved in terms of elliptic functions, and the similarity solution is solved in terms of the Painlev\'e IV transcendent.
Since a number of newly proposed integrable `nonlocal' NLS variants (the $\mathcal{PT}$-symmetric nonlocal NLS, the reverse time NLS, and the reverse space-time NLS) are derivable as specific cases of this system, a consequence is that the nonlocal Painlev\'e type ODEs obtained from these nonlocal variants all reduce to previously known local equations.

\end{abstract}

\section{Introduction}\label{sec:intro}

Integrable nonlinear partial differential equations (PDEs) play an important role in the study of nonlinear wave propagation, and have been studied extensively \cite{AblowitzSolitonsBook,ablowitz2006solitons,novikov1984theory,yang2010nonlinear}. 
Famous examples include the Korteweg-deVries equation, the sine-Gordon equation, and the nonlinear Schrödinger
equation (NLS)
\begin{equation}\label{eqn:NLS}
        iq_t(x,t)=q_{xx}(x,t)-2\sigma q(x,t)^2 q^*(x,t),\quad \sigma=\pm 1,
\end{equation}
where $q$ is a complex valued function of $x$ and $t$, and $^*$ denotes complex conjugation.
Here $\sigma=-1$ corresponds to a focusing nonlinearity, and $\sigma=1$ corresponds to a defocusing nonlinearity. 
Many of these equations constitute fundamental models for a wide range of physical phenomena - for instance, the NLS is used to model wave propagation in nonlinear optical fibres and waveguides \cite{ablowitz2004discrete}, small-amplitude waves on the surface of deep inviscid water, Langmuir waves in hot plasmas, \cite{encyclopedia_nonlin_nls}, and the mean-field regime of Bose-Einstein condensates \cite{dalfovo1999theory}.

Recently, there has been significant interest in newly discovered `nonlocal' variants of previously studied integrable PDEs.
In 2013, an integrable nonlocal NLS (NNLS) was proposed:
\begin{equation}\label{eqn:NNLS}
        iq_t(x,t)=q_{xx}(x,t)-2\sigma q(x,t)^2 q^*(-x,t),\quad \sigma=\pm 1,
\end{equation}
where the `nonlocality' refers to the simultaneous dependence on $x$ and $-x$ \cite{Ablowitz2013}.
This inspired the discovery of many other integrable nonlocal equations \cite{Ablowitz2017,ji2017nonlocal}.
Among these, two further integrable NLS variants that are nonlocal in time were recovered:
the reverse space-time NLS (RST NLS)
\begin{align}\label{eqn:RSTNLS}
    iq_t(x,t)=q_{xx}(x,t)-2\sigma q(x,t)^2 q(-x,-t),\quad \sigma=\pm 1,
\end{align}
and the reverse time NLS (RT NLS)
\begin{align}\label{eqn:RTNLS}
    iq_t(x,t)=q_{xx}(x,t)-2\sigma q(x,t)^2 q(x,-t),\quad \sigma=\pm 1.
\end{align}

Each of these four equations are obtainable via the integrable AKNS system \cite{ablowitz2006solitons}
\begin{subequations}\label{eqn:master_NLS_system}
\begin{align}
    iq_t&=q_{xx}-2q^2 r \label{eqn:system_q},\\
    -ir_t&=r_{xx}-2r^2 q \label{eqn:system_r},
\end{align}
\end{subequations}
which establishes their integrability, since the same inverse scattering procedure will broadly apply to all of them, albeit with new symmetry relationships imposed \cite{Ablowitz2017,yang2019general}.
Under the symmetry reduction 
\begin{equation}\label{eqn:symred_NLS}
    r(x,t)=\sigma q^*(x,t),  
\end{equation}
the system \eqref{eqn:master_NLS_system} becomes equivalent to the classic NLS for $q$.
The nonlocal variants \eqref{eqn:NNLS}, \eqref{eqn:RSTNLS}, \eqref{eqn:RTNLS} are obtained in a similar manner from \eqref{eqn:master_NLS_system} via the reductions
\begin{align}
    \text{NNLS}: && r(x,t)&=\sigma q^*(-x,t),\label{eqn:symred_NNLS}\\
    \text{RST NLS}: && r(x,t)&=\sigma q(-x,-t),\label{eqn:symred_RSTNLS}\\
    \text{RT NLS}: && r(x,t)&=\sigma q(x,-t).\label{eqn:symred_RTNLS}
\end{align}
Notably, each of NLS variants \eqref{eqn:NLS}-\eqref{eqn:RTNLS} can thus be viewed as a specific solution of the system \eqref{eqn:master_NLS_system}.

The reverse space NNLS has been intensely studied since its introduction \cite{wen2016dynamics,huang2016soliton,gerdjikov2017complete,gurses2018nonlocal}.
In particular, an interesting property is that it exhibits so-called `parity-time ($\mathcal{PT}$) symmetry'.
That is, it can be viewed as a quantum mechanical (time-dependent) Schr\"odinger equation 
\begin{align}
    iq_t(x,t)=q_{xx}(x,t)+V(x,t)q(x,t),\\
    V(x,t)=-2\sigma q(x,t)q^*(-x,t),
\end{align}
where the self-induced potential $V(x,t)$ satisfies the $\mathcal{PT}$-symmetry condition $V^*(-x,t)=V(x,t)$ \cite{Ablowitz2013}.
This gives a connection to the field of $\mathcal{PT}$-symmetric quantum mechanics, $\mathcal{PT}$-symmetric optics, and other $\mathcal{PT}$-symmetric physical applications which are currently the subject of much research activity \cite{Bender_intro_PT,el2018non,konotop2016nonlinear}.

The general N-soliton behaviour of each of the three nonlocal NLS variants was investigated recently in \cite{yang2019general}, which demonstrated new soliton behaviours for these equations, arising from novel configurations of eigenvalues in the spectral plane.
The relation between nonlocal and local versions of integrable equations was studied in \cite{yang2018transformations}, which demonstrated that in some cases it is possible to transform from the nonlocal to local case and vice versa.
In terms of applications, the NNLS \eqref{eqn:NNLS} was derived in a physical application of magnetics \cite{gadzhimuradov2016towards}, and more generally, nonlocal space/time couplings in integrable equations may be used to model scenarios involving correlated events \cite{lou2017alice}.

Another interesting aspect of integrable nonlinear PDEs is their relation to nonlinear ordinary differential equations (ODEs). 
It is well-known that reductions of integrable nonlinear PDEs yield (possibly after a transformation) ODEs of Painlev\'e type  - which is to say the solutions of these ODEs do not have any movable singularities that are worse than poles \cite{kruskal1992painleve}.
Notable among Painlev\'e type ODEs are the Painlev\'e equations - a class of nonlinear second-order equations that define special functions called the Painlev\'e transcendents.
These functions have many interesting analytic properties, and arise in a broad range of applications, like in the solutions of integrable PDEs such as the NLS, in statistical physics, random matrices, and quantum field theory \cite{painlevechapter, conte2012painleve,conte2008painleve}.
Whilst extensive efforts have been made to classify Painlev\'e type ODEs, with complete results obtained in certain cases, classification in general remains an open problem \cite{clarkson2019openNEW}.

In the case of the original NLS, it is known that the stationary solution $q(x,t)=e^{i\lambda t}f(x)$ yields the ODE
\begin{equation}\label{eqn:NLS_stat_ODE}
    -\lambda f=f''-2\sigma f^2f^*
\end{equation}
which is solved in terms of elliptic functions (the ODEs that define elliptic functions are of Painlev\'e type).
Further, the similarity solution $q(x,t)=\frac{1}{\sqrt{2t}}Q(z)$, $z=\frac{x}{\sqrt{2t}}$
leads to 
\begin{equation}\label{eqn:NLS_sim_ODE}
    -iQ-izQ''=Q''-2\sigma Q^2 Q^*
\end{equation}
which is solved in terms of Painlev\'e IV \cite{Boiti1980, Can1991}.

It was noted in \cite{Ablowitz2017} that taking analogous reductions for the nonlocal variants of NLS yield new Painlev\'e type equations that are nonlocal analogues of these previously known ones.
In \cite{xu2019general}, the general stationary solution of the NNLS was investigated and reduced down to an elliptic function.
However, to our best knowledge, the other reductions have not been studied.

In this brief communication, we deal with the AKNS system \eqref{eqn:master_NLS_system} in general and via integrating and transforming, obtain a superset of solutions for the 1) stationary solution case and 2) the similarity solution case.
This solution set will also form a superset of the stationary and similarity solutions for each of the four NLS variants.
We demonstrate that the general stationary case reduces to an equation solvable by known elliptic functions, and the general similarity solution reduces to Painlev\'e IV.
That is, the broad outcomes that occur for the reductions of the standard NLS also hold for the other three nonlocal variants.
Further, the nonlocal Painlev\'e type ODEs we obtain are solved in terms of previously known functions and so do not define any truly new functions.

\section{General stationary solution}

Consider the stationary solution
\begin{equation}
    q(x,t)=e^{i\lambda t}f(x),
\end{equation}
where $f(x)$ is a function $\mathbb{R}\to\mathbb{C}$, and $\lambda\in\mathbb{R}$ is a constant.
Then $q_t=i\lambda q$, and for each of the four symmetry reductions \eqref{eqn:symred_NLS} to \eqref{eqn:symred_RTNLS}, we will have $r_t=-i\lambda r$, which can be seen from
\begin{subequations}
\begin{align}
    &\text{NLS:}
    &r(x,t)&=\sigma q^*(x,t)=\sigma e^{-i\lambda t}f^*(x)\label{eqn:stat_rf1},\\
    &\text{NNLS:}
    &r(x,t)&=\sigma q^*(-x,t)=\sigma e^{-i\lambda t}f^*(-x)\label{eqn:stat_rf2},\\
    &\text{RST NLS:}
    &r(x,t)&=\sigma q(-x,-t)=\sigma e^{-i\lambda t}f(-x)\label{eqn:stat_rf3},\\
    &\text{RT NLS:}
    &r(x,t)&=\sigma q(x,-t)=\sigma e^{-i\lambda t}f(x)\label{eqn:stat_rf4}.
\end{align}
\end{subequations}
Further, in each case \eqref{eqn:stat_rf1} to \eqref{eqn:stat_rf4}, we can write $r(x,t):=\sigma e^{-i\lambda t}g(x)$, where $g(x)$ is related to $f(x)$ through some combination of complex conjugation and the operation $x\to -x$.
For instance, in the case of the NNLS reduction \eqref{eqn:stat_rf2} we have $g(x)=f^*(-x)$.

Hence, without loss of generality we can simplify the AKNS system \eqref{eqn:master_NLS_system} to the ODE system
\begin{subequations}\label{eqn:stat_sys}
\begin{align}
    -\lambda f&=f''-2\sigma f^2g, \label{eqn:stat_sys_1}\\
    -\lambda g&=g''-2\sigma g^2 f, \label{eqn:stat_sys_2}
\end{align}
\end{subequations}
which incorporates the stationary solutions of all four NLS variants as specific cases.
In particular, for the nonlocal symmetry reductions, taking the stationary solution yields new Painlev\'e type equations which are nonlocal analogues of \eqref{eqn:NLS_stat_ODE} - i.e. equation \eqref{eqn:stat_sys_1} gives:
\begin{subequations}
\begin{align}
    &\text{NNLS:} &-\lambda f(x)=f''(x)-2\sigma f(x)^2f^*(-x),\label{eqn:PT_elliptic}\\
    &\text{RST NLS:} &-\lambda f(x)=f''(x)-2\sigma f(x)^2f(-x). \label{eqn:RST_elliptic}
\end{align}
\end{subequations}
We observe that \eqref{eqn:PT_elliptic} is the time-independent quantum mechanical Schr\"odinger equation with the $\mathcal{PT}$-symmetric self-induced potential $-2\sigma f(x)f^*(-x)$.

From here, we will neglect the relationship between $f$ and $g$, and solve for the ODE system \eqref{eqn:stat_sys} in general.
We use \eqref{eqn:stat_sys_1} to express $g$ in terms of $f$ and $f''$.
Differentiating twice yields an expression for $g''$ in terms of $f$ and derivatives of $f$.
Thus, we can replace all instances of $g$ and $g''$ in \eqref{eqn:stat_sys_2} to obtain a fourth-order ODE in $f$
\begin{align}\label{eqn:nonlocal_stat_ODE}
    f^{(4)} f^2-2 \lambda  f^2 f''-3 f f''^2+2 \lambda  f
   f'^2-4 f^{(3)} f f'+6 f'^2 f''=0.
\end{align}
Noting the homogeneity of terms, we use the well-known substitution
\begin{equation}
    v(x):=\frac{f'(x)}{f(x)}
\end{equation}
which is valid for $f$ not identically 0, to obtain an ODE of reduced order in the new function $v$
\begin{equation}
    v^{(3)}-2 \lambda  v'-6 v^2 v'=0.
\end{equation}
This can be directly integrated to obtain
\begin{equation}
    v''-2v^3-2\lambda v-a=0.
\end{equation}
where $a\in\mathbb{C}$ is a constant of integration.
Multiplying by $v'$, it can be integrated again to 
\begin{align}
    v'^2=v^4+2\lambda v^2+av+b,
\end{align}
with $a$, $b\in\mathbb{C}$ constants of integration.
The general solution to this ODE is an elliptic function, which can be written as an elliptic integral of the first kind \cite{busam2009complex}.

\section{General similarity solution}

Consider the similarity solution
\begin{align}\label{eqn:NLS_system_sim_soln}
    q(x,t)=\frac{1}{(2t)^{1/2}}Q\left(\frac{x}{(2t)^{1/2}}\right), &&
    r(x,t)=\frac{1}{(2t)^{1/2}}R\left(\frac{x}{(2t)^{1/2}}\right).
\end{align}
Analogously to the previous section, consider what happens for each specific reduction.
The $\mathcal{PT}$-symmetric case behaves nicely:
\begin{subequations}
\begin{align}
    &\text{NLS} &&r(x,t)=\sigma q^*(x,t) \iff R(z)=\sigma Q^*(z),\\
    &\text{NNLS} &&r(x,t)=\sigma q^*(-x,t) \iff R(z)=\sigma Q^*(-z).
\end{align}
\end{subequations}
However, when time reversal is involved, we see
\begin{subequations}
\begin{align}
    \text{RT NLS} &&
    r(x,t)=\sigma q(x,-t)
    &\iff \frac{1}{\sqrt{2t}}R\left(\frac{x}{\sqrt{2t}}\right)
    =\frac{\sigma}{\sqrt{-2t}}Q\left(\frac{x}{\sqrt{-2t}}\right)\\
    \text{RST NLS}, &&
    r(x,t)=\sigma q(-x,-t)
    &\iff \frac{1}{\sqrt{2t}}R\left(\frac{x}{\sqrt{2t}}\right)
    =\frac{\sigma}{\sqrt{-2t}}Q\left(-\frac{x}{\sqrt{-2t}}\right), 
\end{align}
\end{subequations}
i.e., there is branch dependence, as noted in \cite{Ablowitz2017}.
Let $\kappa=\frac{1}{\sqrt{-1}}=\pm i$.
Then the above two conditions become
\begin{subequations}
\begin{align}
    &\text{RT NLS}
    &R(z)&=\sigma\kappa Q(\kappa z),\\
    &\text{RST NLS}
    &R(z)&=\sigma \kappa Q(-\kappa z).
\end{align}
\end{subequations}
Substituting in, the AKNS system \eqref{eqn:master_NLS_system} reduces to the ODE system
\begin{subequations}\label{eqn:sim_sys}
\begin{align}
    -iQ-izQ' &=Q''-2 Q^2R,\label{eqn:sim_sys_1}\\
    iR+izR' &=R''-2 R^2Q. \label{eqn:sim_sys_2}
\end{align}
\end{subequations}
In the cases of the three nonlocal symmetry reductions, we find nonlocal Painlev\'e equations
\begin{subequations}
\begin{align}
    &\text{NNLS} &&-iQ(z)-izQ'(z)=Q''(z)-2\sigma Q(z)^2 Q^*(-z),\label{eqn:PT_painleve}\\
    &\text{RT NLS} &&-iQ(z)-izQ'(z)=Q''(z)-2\sigma\kappa Q(z)^2 Q(\kappa z),\label{eqn:RT_painleve}\\
    &\text{RST NLS} &&-iQ(z)-izQ'(z)=Q''(z)-2\sigma\kappa Q(z)^2 Q(-\kappa z).\label{eqn:RST_painleve}
\end{align}
\end{subequations}
The first one can be thought of as a $\mathcal{PT}$-symmetric Painlev\'e equation \cite{Ablowitz2017}.

Again, we consider $Q$, $R$ to be uncoupled and solve for \eqref{eqn:sim_sys} in general.
Equation \eqref{eqn:sim_sys_1} can be used to express $R$ in terms of $Q$, $Q'$, and $Q''$.
This can then be used to replace all instances of $R$ and its derivatives in \eqref{eqn:sim_sys_2} to obtain a fourth-order ODE in $Q$
\begin{align}
    -&6 i z Q'^3-6 Q'^2 Q''-Q^2 \left(Q^{(4)}+\left(z^2-2
   i\right) Q''+3 z Q'\right)\nonumber\\
   &+Q \left(3 Q''^2+\left(z^2+2
   i\right) Q'^2+Q' \left(4 Q^{(3)}+6 i z Q''\right)\right)-2
   Q^3=0.
\end{align}
Noting again the homogeneity, substitute
\begin{equation}
    V(z):=\frac{Q'(z)}{Q(z)}
\end{equation}
to obtain
\begin{equation}\label{eqn:w'''}
    V^{(3)}+z^2 V'-6 i z V V'-6 V^2 V'-2 i V'+3 z V-4
   i V^2+2=0.
\end{equation}
Taking
\begin{align}
    V(z)=Y(z)-\frac{i}{2}z
\end{align}
transforms \eqref{eqn:w'''} to
\begin{equation}\label{eqn:A.thirdorder}
    Y^{(3)}=6Y^2Y'+\left(\frac{z^2}{2}+2i\right)\left(Y'+\frac{i}{2}\right)+iY^2+zY
\end{equation}
which is sometimes known as Chazy VIII.a (see section 6.8 of \cite{Cosgrove2000}) with parameters $\alpha=\frac{i}{2}$, $\beta=0$, $\gamma=2i$.
The Chazy VIII.a equation has a general solution in terms of Painlev\'{e} IV: a final transformation
\begin{align}
    Y(z)
    = i\sqrt{\alpha}w(\zeta)-\alpha z
    =\frac{1}{2}(-1+i)w\left(\frac{1}{2}(-1+i)z\right)-\frac{i}{2}z
\end{align}
takes \eqref{eqn:A.thirdorder} to the canonical Painlev\'{e} IV 
\begin{equation}
    w''=\frac{1}{2w}w'^2+\frac{3}{2}w^3+4zw^2+2(z^2-A)w+\frac{B}{w}
\end{equation}
with parameters $A=1$, and $B$ arbitrary.

\section{Conclusion}
Two simple ODE reductions of the AKNS system associated with NLS type equations were reduced to well known Painlev\'e type ODEs through a series of integrations.
This includes as specific cases a number of nonlocal and/or $\mathcal{PT}$-symmetric Painlev\'e type ODEs that come from reductions of new nonlocal NLS variants.

The simple method we used to deal with ODEs with nonlocal dependency seems like
it could be applicable in other cases. That is to say, a future step would be repeating the analysis for
some other integrable PDE with ODE reductions known to be solved in terms of Painlevé type
functions. 
One would take the nonlocal version of this PDE, apply an ODE reduction to get another nonlocal Painlevé type ODE, use symmetry to generate a system of equations, and finally recover a single local ODE which can be integrated repeatedly.

% \iffalse%%%
\section*{Acknowledgements}
We would like to thank Professor Nalini Joshi of the University of Sydney for her supervision of this project.
% \fi %%%

\sloppy
\printbibliography

@article{Bender_intro_PT,
author = {Bender, Carl M.    },
title = {Introduction to $\mathcal{PT}$-symmetric quantum theory},
journal = {Contemporary Physics},
volume = {46},
number = {4},
pages = {277-292},
year  = {2005},
publisher = {Taylor & Francis},
doi = {10.1080/00107500072632},

URL = { 
        https://doi.org/10.1080/00107500072632
    
},
eprint = { 
        https://doi.org/10.1080/00107500072632
    
}

}

@article{Ablowitz2013,
author = {Ablowitz, Mark J. and Musslimani, Ziad H.},
doi = {10.1103/PhysRevLett.110.064105},
issn = {00319007},
journal = {Phys. Rev. Lett.},
number = {6},
pages = {1--5},
title = {{Integrable nonlocal nonlinear Schr{\"{o}}dinger equation}},
volume = {110},
year = {2013}
}

@article{Can1991,
author = {Can, M},
doi = {10.1007/BF02827336},
issn = {03693554},
journal = {Il Nuovo Cimento},
number = {2},
pages = {205--207},
title = {{On the relation between nonlinear Schr{\"{o}}dinger equation and Painlev{\'{e}} IV equation}},
volume = {106},
year = {1991}
}

@article{yang2018transformations,
  title={Transformations between nonlocal and local integrable equations},
  author={Yang, Bo and Yang, Jianke},
  journal={Stud. Appl. Math.},
  volume={140},
  number={2},
  pages={178--201},
  year={2018},
  publisher={Wiley Online Library}
}

@Article{Boiti1980,
author="Boiti, M.
and Pempinelli, F.",
title="Nonlinear {S}chr{\"o}dinger equation, {B}{\"a}cklund transformations and {P}ainlev{\'e} transcendents",
journal="Il Nuovo Cimento B (1971-1996)",
year="1980",
day="01",
volume="59",
number="1",
pages="40--58",
issn="1826-9877",
doi="10.1007/BF02739045",
url="https://doi.org/10.1007/BF02739045"
}

@book{AblowitzSolitonsBook,
author = "Ablowitz, Mark J.",
address = "Cambridge",
booktitle = "Solitons, Nonlinear Evolution Equations and Inverse Scattering",
lccn = "92159759",
publisher = "Cambridge Univ. Press",
series = "London Math. Soc. Lecture Note Ser.",
volume = "149",
title = "Solitons, Nonlinear Evolution Equations and Inverse Scattering ",
year = "1991",
}

@article{Ablowitz2017,
archivePrefix = {arXiv},
arxivId = {1610.02594},
author = {Ablowitz, Mark J. and Musslimani, Ziad H.},
doi = {10.1111/sapm.12153},
eprint = {1610.02594},
issn = {14679590},
journal = {Stud. Appl. Math.},
number = {1},
pages = {7--59},
title = {Integrable nonlocal nonlinear equations},
volume = {139},
year = {2017}
}

@article{Cosgrove2000,
author = {Cosgrove, Christopher M.},
doi = {10.1111/1467-9590.00134},
issn = {00222526},
journal = {Stud. Appl. Math.},
number = {3},
pages = {171--228},
title = {{Chazy classes IX-XI of third-order differential equations}},
volume = {104},
year = {2000}
}

@article{gadzhimuradov2016towards,
  title={Towards a gauge-equivalent magnetic structure of the nonlocal nonlinear {S}chr{\"o}dinger equation},
  author={Gadzhimuradov, TA and Agalarov, AM},
  journal={Phys. Rev. A},
  volume={93},
  number={6},
  pages={062124},
  year={2016},
  publisher={APS}
}

@article{konotop2016nonlinear,
  title={Nonlinear Waves in {PT}-symmetric Systems},
  author={Konotop, Vladimir V and Yang, Jianke and Zezyulin, Dmitry A},
  journal={Rev. Modern Phys.},
  volume={88},
  number={3},
  pages={035002},
  year={2016},
  publisher={APS}
}

@book{yang2010nonlinear,
  title={Nonlinear Waves in Integrable and Nonintegrable Systems},
  author={Yang, Jianke},
  volume={16},
  year={2010},
  publisher={SIAM}
}

@article{yang2019general,
  title={General N-solitons and their dynamics in several nonlocal nonlinear {S}chr{\"o}dinger equations},
  author={Yang, Jianke},
  journal={Phys. Lett. A},
  volume={383},
  number={4},
  pages={328--337},
  year={2019},
  publisher={Elsevier}
}

@article{el2018non,
  title={Non-{H}ermitian physics and {PT} symmetry},
  author={El-Ganainy, Ramy and Makris, Konstantinos G and Khajavikhan, Mercedeh and Musslimani, Ziad H and Rotter, Stefan and Christodoulides, Demetrios N},
  journal={Nature Physics},
  volume={14},
  number={1},
  pages={11--19},
  year={2018},
  publisher={Nature Publishing Group}
}

@incollection{painlevechapter,
  author       = {Peter A. Clarkson}, 
  title        = {Painlev\'{e} equations - nonlinear special functions},
  booktitle    = {Orthogonal Polynomials and Special Functions: Computation and Applications},
  publisher    = {Springer},
  year         = 2006,
  editor       = {F. Marcell\'{a}n, W. Van Assche},
  volume       = 1883,
  series       = {Lecture Notes in Math.},
  pages        = {331-411},
}

@article{lou2017alice,
  title={Alice-{B}ob physics: coherent solutions of nonlocal {K}d{V} systems},
  author={Lou, SY and Huang, Fei},
  journal={Scientific reports},
  volume={7},
  number={1},
  pages={1--11},
  year={2017},
  publisher={Nature Publishing Group}
}

@incollection{encyclopedia_nonlin_nls,
  author       = {Boris Malomed}, 
  title        = {Nonlinear {S}chr\"{o}dinger equations},
  booktitle    = {Encyclopedia of Nonlinear Science},
  publisher    = {Routledge},
  year         = 2006,
  editor       = {Scott Alwyn},
  pages        = {639-643},
}

@article{ji2017nonlocal,
  title={On a nonlocal modified {K}orteweg-de {V}ries equation: integrability, {D}arboux transformation and soliton solutions},
  author={Ji, Jia Liang and Zhu, Zuo Nong},
  journal={Commun. Nonlinear Sci. Numer. Simul.},
  volume={42},
  pages={699--708},
  year={2017},
  publisher={Elsevier}
}

@book{novikov1984theory,
  title={Theory of Solitons: the Inverse Scattering Method},
  author={Novikov, S and Manakov, SV and Pitaevskii, LP and Zakharov, Vladimir E},
  year={1984},
  publisher={Springer}
}

@book{ablowitz2006solitons,
  title={Solitons and the Inverse Scattering Transform},
  author={Ablowitz, Mark J and Segur, Harvey},
  volume={4},
  year={2006},
  publisher={SIAM}
}

@book{conte2012painleve,
  title={The Painlev{\'e} Property: One Century Later},
  author={Conte, Robert},
  year={2012},
  publisher={Springer}
}

@book{conte2008painleve,
  title={The Painlev{\'e} Handbook},
  author={Conte, Robert and Musette, Micheline},
  year={2008},
  publisher={Springer}
}

@article{dalfovo1999theory,
  title={Theory of {B}ose-{E}instein condensation in trapped gases},
  author={Dalfovo, Franco and Giorgini, Stefano and Pitaevskii, Lev P and Stringari, Sandro},
  journal={Rev. Modern Phys.},
  volume={71},
  number={3},
  pages={463},
  year={1999},
  publisher={APS}
}

@book{ablowitz2004discrete,
  title={Discrete and Continuous Nonlinear Schr{\"o}dinger Systems},
  author={Ablowitz, Mark J and Prinari, B and Trubatch, AD},
  series = "London Math. Soc. Lecture Note Ser.",
  volume={302},
  year={2004},
  publisher={Cambridge Univ. Press}
}

@article{xu2019general,
  title={General stationary solutions of the nonlocal nonlinear {S}chr{\"o}dinger equation and their relevance to the {PT}-symmetric system},
  author={Xu, Tao and Chen, Yang and Li, Min and Meng, De Xin},
  journal={Chaos},
  volume={29},
  number={12},
  pages={123124},
  year={2019},
  publisher={AIP Publishing LLC}
}

@book{busam2009complex,
  title={Complex Analysis},
  author={Busam, Rolf and Freitag, Eberhard},
  year={2009},
  publisher={Springer}
}

@article{wen2016dynamics,
  title={Dynamics of higher-order rational solitons for the nonlocal nonlinear {S}chr{\"o}dinger equation with the self-induced parity-time-symmetric potential},
  author={Wen, Xiao Yong and Yan, Zhenya and Yang, Yunqing},
  journal={Chaos},
  volume={26},
  number={6},
  pages={063123},
  year={2016},
  publisher={AIP Publishing LLC}
}

@article{huang2016soliton,
  title={Soliton solutions for the nonlocal nonlinear {S}chr{\"o}dinger equation},
  author={Huang, Xin and Ling, Liming},
  journal={The European Physical Journal Plus},
  volume={131},
  number={5},
  pages={1--11},
  year={2016},
  publisher={Springer}
}

@article{gerdjikov2017complete,
  title={Complete integrability of nonlocal nonlinear {S}chr{\"o}dinger equation},
  author={Gerdjikov, VS and Saxena, Avadh},
  journal={J. Math. Phys.},
  volume={58},
  number={1},
  pages={013502},
  year={2017},
  publisher={AIP Publishing LLC}
}

@article{gurses2018nonlocal,
  title={Nonlocal nonlinear {S}chr{\"o}dinger equations and their soliton solutions},
  author={G{\"u}rses, Metin and Pekcan, Asl{\i}},
  journal={J. Math. Phys.},
  volume={59},
  number={5},
  pages={051501},
  year={2018},
  publisher={AIP Publishing LLC}
}

@article{kruskal1992painleve,
  title={The {P}ainlev{\'e}-{K}owalevski and poly-{P}ainlev{\'e} tests for integrability},
  author={Kruskal, Martin D and Clarkson, Peter A},
  journal={Stud. Appl. Math.},
  volume={86},
  number={2},
  pages={87--165},
  year={1992},
  publisher={Wiley Online Library}
}

@article{clarkson2019openNEW,
  title={Open problems for {P}ainlev\'e equations},
  author={Clarkson, Peter A},
  journal={SIGMA Symmetry Integrability Geom. Methods Appl.},
  volume={15},
  pages={006},
  year={2019},
  publisher={SIGMA. Symmetry, Integrability and Geometry: Methods and Applications}
}
\fussy

\end{document}